# Birkhoff's Completeness Theorem for Multi-Sorted Algebras Formalized in Agda

Andreas Abel
Department of Computer Science, Gothenburg University, Sweden

3 June 2021

This document provides a formal proof of Birkhoff's completeness theorem for multi-sorted algebras which states that any equational entailment valid in all models is also provable in the equational theory. More precisely, if a certain equation is valid in all models that validate a fixed set of equations, then this equation is derivable from that set using the proof rules for a congruence.

The proof has been formalized in Agda version 2.6.2 with the Agda Standard Library version 1.7 and this document reproduces the commented Agda code.

## 1 Introduction

Birkhoff's completeness theorem [1935] has been formalized in type theory before, even in Agda [Gunther et al., 2017, Thm .3.1]. Our formalization makes the following decisions:

1. We use indexed containers [Altenkirch et al., 2015] aka Peterson-Synek (interaction) trees. Given a set $S$ of sort symbols, a signature over $S$ is an indexed endo-container, which has three components:

   a) Per sort $s : S$, a set $O_s$ of operator symbols. (In the container terminology, these are called *shapes* for index $s$, and in the interaction tree terminology, *commands* for state $s$.)

   b) Per operator $o : O_s$, a set $A_o$, the arity of operator $o$. The arity is the index set for the arguments of the operator, which are then given by a function over $A_o$. (In the other terminologies, these are the *positions* or *responses*, resp.)

   c) Per argument index $i : A_o$, a sort $s_i : S$ which denotes the sort of the $i$th argument of operator $o$. (In the interaction tree terminology, this is the *next* state.)

   Closed terms of a multi-sorted algebra (aka first-order terms) are then concrete interaction trees, i.e., elements of the indexed $W$-type pertaining to the container.

   Note that all the "set"s we mentioned above come with a size, see next point.

2. Universe-polymorphic: As we are working in a predicative and constructive meta-theory, we have to be aware of the size (i.e., inaccessible cardinality) of the sets. Our formalization is universe-polymorphic to ensure good generality, resting on the universe-polymorphic Agda Standard Library.

   In particular, there is no such thing as "all models"; rather we can only quantify over models of a certain maximum size. The completeness theorem consequently does not require validity of an entailment in all models, but only in all models of a certain size, which is given by the size of the generic model, i.e., the term model. The size of the term model in turn is determined by the size of the signature of the multi-sorted algebra.

3. Open terms (with free variables) are obtained as the *free monad* over the container. Concretely, we make a new container that has additional nullary operator symbols, which stand for the variables. Terms are intrinsically typed, i.e., the set of terms is actually a family of sets indexed by a sort and a context of sorted variables in scope.

4. No lists: We have no finiteness restrictions whatsoever, neither the number of operators need to be finite, nor the number of arguments of an operator, nor the set of variables that are in scope of a term. (Note however, since terms are finite trees, they can actually mention only a finite number of variables from the possibly infinite supply.)

## 2 Preliminaries

We import library content for indexed containers, standard types, and setoids.

```
open import Level

open import Data.Container.Indexed.Core         using (Container; ⟦_⟧; _◃_/_)
open import Data.Container.Indexed.FreeMonad    using (_⋆C_)
open import Data.W.Indexed                      using (W; sup)

open import Data.Product                        using (Σ; _×_; _,_; Σ-syntax); open Σ
open import Data.Sum                            using (_⊎_; inj₁; inj₂; [_,_])
open import Data.Empty.Polymorphic              using (⊥; ⊥-elim)

open import Function                            using (_∘_)
open import Function.Bundles                    using (Func)

open import Relation.Binary                     using (Setoid; IsEquivalence)
open import Relation.Binary.PropositionalEquality using (_≡_; refl)
open import Relation.Unary                      using (Pred)

import Relation.Binary.Reasoning.Setoid as SetoidReasoning

open Setoid using (Carrier; _≈_; isEquivalence)
open Func renaming (f to apply)
```

Letter $\ell$ denotes universe levels.

```
variable
  ℓ ℓ' ℓˢ ℓᵒ ℓᵃ ℓᵐ ℓᵉ ℓⁱ : Level
  I : Set ℓⁱ
  S : Set ℓˢ
```

The interpretation of a container (Op ◁ Ar / sort) is

⟦ Op ◁ Ar / sort ⟧ X s = Σ[ o ∈ Op s ] ((i : Ar o) → X (sort o i))

which contains pairs consisting of an operator $o$ and its collection of arguments. The least fixed point of (X ↦ ⟦ C ⟧ X) is the indexed W-type given by C, and it contains closed first-order terms of the multi-sorted algebra C.

We need to interpret indexed containers on Setoids. This definition is missing from the standard library v1.7. It equips the sets (⟦ C ⟧ X s) with an equivalence relation induced by the one of the family $X$. The definition of ⟦_⟧$s$ can be stated for heterogeneous index containers where we distinguish input and output sorts $I$ and $S$.

```
⟦_⟧s : (C : Container I S ℓᵒ ℓᵃ) (ξ : I → Setoid ℓᵐ ℓᵉ) → S → Setoid _ _

⟦ C ⟧s ξ s .Carrier =
  ⟦ C ⟧ (Carrier ∘ ξ) s

⟦ Op ◁ Ar / sort ⟧s ξ s ._≈_ (op , args) (op' , args') =
  Σ[ eq ∈ op ≡ op' ] EqArgs eq args args'
  where
  EqArgs : (eq : op ≡ op')
           (args  : (i : Ar op)  → ξ (sort _ i) .Carrier)
           (args' : (i : Ar op') → ξ (sort _ i) .Carrier)
         → Set _
  EqArgs refl args args' = (i : Ar op) → ξ (sort _ i) ._≈_ (args i) (args' i)

⟦ Op ◁ Ar / sort ⟧s ξ s .isEquivalence .IsEquivalence.refl
                         = refl , λ i → Setoid.refl   (ξ (sort _ i))
⟦ Op ◁ Ar / sort ⟧s ξ s .isEquivalence .IsEquivalence.sym   (refl , g)
                         = refl , λ i → Setoid.sym   (ξ (sort _ i)) (g i)
⟦ Op ◁ Ar / sort ⟧s ξ s .isEquivalence .IsEquivalence.trans (refl , g) (refl , h)
                         = refl , λ i → Setoid.trans (ξ (sort _ i)) (g i) (h i)
```

## 3 Multi-sorted algebras

A multi-sorted algebra is an indexed container.

- Sorts are indexes.
- Operators are commands/shapes.
- Arities/argument are responses/positions.

Closed terms (initial model) are given by the W type for a container, renamed to $\mu$ here (for least fixed-point).

It is convenient to name the concept of signature, i.e. (Sort, Ops)

```
record Signature (ℓˢ ℓᵒ ℓᵃ : Level) : Set (suc (ℓˢ ⊔ ℓᵒ ⊔ ℓᵃ)) where
  field
    Sort : Set ℓˢ
    Ops : Container Sort Sort ℓᵒ ℓᵃ
```

We assume a fixed signature.

```
module _ (Sig : Signature ℓˢ ℓᵒ ℓᵃ) where
  open Signature Sig
  open Container Ops renaming
    ( Command to Op
    ; Response  to Arity
    ; next      to sort
    )
```

We let letter $s$ range over sorts and $op$ over operators.

```
variable
  s s'     : Sort
  op op'   : Op s
```

## 3.1 Models

A model is given by an interpretation (Den $s$) for each sort $s$ plus an interpretation (den $o$) for each operator $o$. A model is also frequently known as an *Algebra* for a signature; but as that terminology is too overloaded, it is avoided here.

```
record SetModel ℓᵐ : Set (ℓˢ ⊔ ℓᵒ ⊔ ℓᵃ ⊔ suc ℓᵐ) where
  field
    Den : Sort → Set ℓᵐ
    den : {s : Sort} → ⟦ Ops ⟧ Den s → Den s
```

The setoid model requires operators to respect equality. The Func record packs a function (apply) with a proof (cong) that the function maps equals to equals.

```
record SetoidModel ℓᵐ ℓᵉ : Set (ℓˢ ⊔ ℓᵒ ⊔ ℓᵃ ⊔ suc (ℓᵐ ⊔ ℓᵉ)) where
  field
    Den : Sort → Setoid ℓᵐ ℓᵉ
    den : {s : Sort} → Func (⟦ Ops ⟧s Den s) (Den s)
```

## 4 Terms

To obtain terms with free variables, we add additional nullary operators, each representing a variable.

These are covered in the standard library FreeMonad module, albeit with the restriction that the operator and variable sets have the same size.

```
Cxt : Set (ℓˢ ⊔ suc ℓᵒ)
Cxt = Sort → Set ℓᵒ

variable
  Γ Δ : Cxt
```

Terms with free variables in Var.

```
module _ (Var : Cxt) where
```

We keep the same sorts, but add a nullary operator for each variable.

```
Ops⁺ : Container Sort Sort ℓᵒ ℓᵃ
Ops⁺ = Ops ⋆C Var
```

Terms with variables are then given by the W-type for the extended container.

```
Tm : Pred Sort _
Tm = W Ops⁺
```

We define nice constructors for variables and operator application via pattern synonyms. Note that the $f$ in constructor var' is a function from the empty set, so it should be uniquely determined. However, Agda's equality is more intensional and will not identify all functions from the empty set. Since we do not make use of the axiom of function extensionality, we sometimes have to consult the extensional equality of the function setoid.

```
pattern _•_ op args = sup (inj₂ op , args)
pattern var′ x f    = sup (inj₁ x , f )
pattern var x       = var′ x _
```

Letter $t$ ranges over terms, and $ts$ over argument vectors.

```
variable
  t t′ t₁ t₂ t₃ : Tm Γ s
  ts ts′        : (i : Arity op) → Tm Γ (sort _ i)
```

### 4.1 Parallel substitutions

A substitution from $\Delta$ to $\Gamma$ holds a term in $\Gamma$ for each variable in $\Delta$.

```
Sub : (Γ Δ : Cxt) → Set _
Sub Γ Δ = ∀{s} (x : Δ s) → Tm Γ s
```

Application of a substitution.

```
_[_] : (t : Tm Δ s) (σ : Sub Γ Δ) → Tm Γ s
(var x )  [ σ ] = σ x
(op • ts) [ σ ] = op • λ i → ts i [ σ ]
```

Letter $\sigma$ ranges over substitutions.

```
variable
  σ σ′ : Sub Γ Δ
```

## 5 Interpretation of terms in a model

Given an algebra $M$ of set-size $\ell^m$ and equality-size $\ell^e$, we define the interpretation of an $s$-sorted term $t$ as element of $M(s)$ according to an environment $\rho$ that maps each variable of sort $s'$ to an element of $M(s')$.

```
module _ {M : SetoidModel ℓᵐ ℓᵉ} where
  open SetoidModel M
```

Equality in $M$'s interpretation of sort $s$.

```
_≃_ : Den s .Carrier → Den s .Carrier → Set _
_≃_ {s = s} = Den s ._≈_
```

An environment for $\Gamma$ maps each variable $x : \Gamma(s)$ to an element of $M(s)$. Equality of environments is defined pointwise.

```
Env : Cxt → Setoid _ _
Env Γ .Carrier    = {s : Sort} (x : Γ s) → Den s .Carrier
Env Γ ._≈_ ρ ρ′ = {s : Sort} (x : Γ s) → ρ x ≃ ρ′ x
Env Γ .isEquivalence .IsEquivalence.refl    {s = s} x = Den s .Setoid.refl
Env Γ .isEquivalence .IsEquivalence.sym     h {s} x = Den s .Setoid.sym (h x)
Env Γ .isEquivalence .IsEquivalence.trans g h {s} x = Den s .Setoid.trans (g x) (h x)
```

Interpretation of terms is iteration on the W-type. The standard library offers 'iter' (on sets), but we need this to be a Func (on setoids).

```
⦅_⦆ : ∀{s} (t : Tm Γ s) → Func (Env Γ) (Den s)
⦅ var x      ⦆ .apply ρ      = ρ x
⦅ var x      ⦆ .cong  ρ=ρ′ = ρ=ρ′ x
⦅ op • args ⦆ .apply ρ      = den .apply (op  , λ i → ⦅ args i ⦆ .apply ρ)
⦅ op • args ⦆ .cong  ρ=ρ′ = den .cong  (refl , λ i → ⦅ args i ⦆ .cong ρ=ρ′)
```

An equality between two terms holds in a model if the two terms are equal under all valuations of their free variables.

```
Equal : ∀ {Γ s} (t t′ : Tm Γ s) → Set _
Equal {Γ} {s} t t′ = ∀ (ρ : Env Γ .Carrier) → ⦅ t ⦆ .apply ρ ≃ ⦅ t′ ⦆ .apply ρ
```

This notion is an equivalence relation.

```
isEquiv : IsEquivalence (Equal {Γ = Γ} {s = s})
isEquiv {s = s} .IsEquivalence.refl  ρ        = Den s .Setoid.refl
isEquiv {s = s} .IsEquivalence.sym e ρ        = Den s .Setoid.sym (e ρ)
isEquiv {s = s} .IsEquivalence.trans e e′ ρ = Den s .Setoid.trans (e ρ) (e′ ρ)
```

### 5.1 Substitution lemma

Evaluation of a substitution gives an environment.

```
⦅_⦆s : Sub Γ Δ → Env Γ .Carrier → Env Δ .Carrier
⦅ σ ⦆s ρ x = ⦅ σ x ⦆ .apply ρ
```

Substitution lemma: ⦅t[σ]⦆$\rho \simeq$ ⦅t⦆⦅$\sigma$⦆$\rho$

```
substitution : (t : Tm Δ s) (σ : Sub Γ Δ) (ρ : Env Γ .Carrier) →
  ⦅ t [ σ ] ⦆ .apply ρ ≃ ⦅ t ⦆ .apply (⦅ σ ⦆s ρ)
substitution (var x)   σ ρ = Den _ .Setoid.refl
substitution (op • ts) σ ρ = den .cong (refl , λ i → substitution (ts i) σ ρ)
```

## 6 Equations

An equation is a pair $t \doteq t'$ of terms of the same sort in the same context.

```
record Eq : Set (ℓˢ ⊔ suc ℓᵒ ⊔ ℓᵃ) where
  constructor _≐_
  field
    {cxt} : Sort → Set ℓᵒ
```

```
{srt} : Sort
lhs   : Tm cxt srt
rhs   : Tm cxt srt
```

Equation $t \doteq t'$ holding in model $M$.

```
_⊧_ : (M : SetoidModel ℓᵐ ℓᵉ) (eq : Eq) → Set _
M ⊧ (t ≐ t′) = Equal {M = M} t t′
```

Sets of equations are presented as collection E : I → Eq for some index set I : Set $\ell^i$.

An entailment/consequence $E \supset t \doteq t'$ is valid if $t \doteq t'$ holds in all models that satisfy equations $E$.

```
module _ {ℓᵐ ℓᵉ} where

  _⊃_ : (E : I → Eq) (eq : Eq) → Set _
  E ⊃ eq = ∀ (M : SetoidModel ℓᵐ ℓᵉ) → (∀ i → M ⊧ E i) → M ⊧ eq
```

## 6.1 Derivations

Equalitional logic allows us to prove entailments via the inference rules for the judgment $E \vdash \Gamma \rhd t \equiv t'$. This could be coined as equational theory over a given set of equations $E$. Relation $E \vdash \Gamma \rhd _ \equiv _$ is the least congruence over the equations $E$.

```
data _⊢_▹_≡_ {I : Set ℓⁱ}
  (E : I → Eq) : (Γ : Cxt) (t t′ : Tm Γ s) → Set (ℓˢ ⊔ suc ℓᵒ ⊔ ℓᵃ ⊔ ℓⁱ) where

  hyp   : ∀ i → let t ≐ t′ = E i in
          E ⊢ _ ▹ t ≡ t′

  base  : ∀ (x : Γ s) {f f′ : (i : ⊥) → Tm _ (⊥−elim i)} →
          E ⊢ Γ ▹ var′ x f ≡ var′ x f′

  app   : (∀ i → E ⊢ Γ ▹ ts i ≡ ts′ i) →
          E ⊢ Γ ▹ (op • ts) ≡ (op • ts′)

  sub   : E ⊢ Δ ▹ t ≡ t′ →
          ∀ (σ : Sub Γ Δ) →
          E ⊢ Γ ▹ (t [ σ ]) ≡ (t′ [ σ ])

  refl  : ∀ (t : Tm Γ s) →
          E ⊢ Γ ▹ t ≡ t

  sym   : E ⊢ Γ ▹ t ≡ t′ →
          E ⊢ Γ ▹ t′ ≡ t

  trans : E ⊢ Γ ▹ t₁ ≡ t₂ →
          E ⊢ Γ ▹ t₂ ≡ t₃ →
          E ⊢ Γ ▹ t₁ ≡ t₃
```

### 6.2 Soundness of the inference rules

We assume a model $M$ that validates all equations in $E$.

```
module Soundness {I : Set ℓⁱ} (E : I → Eq) (M : SetoidModel ℓᵐ ℓᵉ)
  (V : ∀ i → M ⊧ E i) where
  open SetoidModel M
```

In any model $M$ that satisfies the equations $E$, derived equality is actual equality.

```
sound : E ⊢ Γ ▹ t ≡ t′ → M ⊧ (t ≐ t′)

sound (hyp i)                    = V i
sound (app {op = op} es) ρ       = den .cong (refl , λ i → sound (es i) ρ)
sound (sub {t = t} {t′ = t′} e σ) ρ = begin
  ⦅ t [ σ ]   ⦆ .apply ρ  ≈⟨ substitution {M = M} t σ ρ ⟩
  ⦅ t         ⦆ .apply ρ′ ≈⟨ sound e ρ′ ⟩
  ⦅ t′        ⦆ .apply ρ′ ≈˘⟨ substitution {M = M} t′ σ ρ ⟩
  ⦅ t′ [ σ ]  ⦆ .apply ρ  ∎
  where
  open SetoidReasoning (Den _)
  ρ′ = ⦅ σ ⦆s ρ

sound (base x {f} {f′})          = isEquiv {M = M} .IsEquivalence.refl {var′ x λ()}

sound (refl t)                   = isEquiv {M = M} .IsEquivalence.refl {t}
sound (sym {t = t} {t′ = t′} e)  = isEquiv {M = M} .IsEquivalence.sym
                                     {x = t} {y = t′} (sound e)
sound (trans {t₁ = t₁} {t₂ = t₂}
             {t₃ = t₃} e e′)     = isEquiv {M = M} .IsEquivalence.trans
                                     {i = t₁} {j = t₂} {k = t₃} (sound e) (sound e′)
```

## 7 Birkhoff's completeness theorem

Birkhoff proved that any equation $t \doteq t'$ is derivable from $E$ when it is valid in all models satisfying $E$. His proof (for single-sorted algebras) is a blue print for many more completeness proofs. They all proceed by constructing a universal model aka term model. In our case, it is terms quotiented by derivable equality $E \vdash \Gamma \rhd _ \equiv _$. It then suffices to prove that this model satisfies all equations in $E$.

### 7.1 Universal model

A term model for $E$ and $\Gamma$ interprets sort $s$ by (Tm Γ s) quotiented by $E \vdash \Gamma \rhd _ \equiv _$.

```
module TermModel {I : Set ℓⁱ} (E : I → Eq) where
  open SetoidModel
```

Tm Γ s quotiented by E⊢Γ▷·≡·.

```
TmSetoid : Cxt → Sort → Setoid _ _
TmSetoid Γ s .Carrier                                = Tm Γ s
TmSetoid Γ s ._≈_                                    = E ⊢ Γ ▷_≡_
TmSetoid Γ s .isEquivalence .IsEquivalence.refl      = refl _
TmSetoid Γ s .isEquivalence .IsEquivalence.sym       = sym
TmSetoid Γ s .isEquivalence .IsEquivalence.trans     = trans
```

The interpretation of an operator is simply the operator. This works because $E \vdash \Gamma \rhd _ \equiv _$ is a congruence.

```
tmInterp : ∀ {Γ s} → Func (⟦ Ops ⟧s (TmSetoid Γ) s) (TmSetoid Γ s)
tmInterp .apply (op , ts) = op • ts
tmInterp .cong (refl , h) = app h
```

The term model per context Γ.

```
M : Cxt → SetoidModel _ _
M Γ .Den = TmSetoid Γ
M Γ .den = tmInterp
```

The identity substitution $\sigma_0$ maps variables to themselves.

```
σ₀ : {Γ : Cxt} → Sub Γ Γ
σ₀ x = var′ x λ()
```

$\sigma_0$ acts indeed as identity.

```
identity : (t : Tm Γ s) → E ⊢ Γ ▷ t [ σ₀ ] ≡ t
identity (var x)    = base x
identity (op • ts) = app λ i → identity (ts i)
```

Evaluation in the term model is substitution $E \vdash \Gamma \rhd ⦅t⦆\sigma \equiv t[\sigma]$. This would even hold "up to the nose" if we had function extensionality.

```
evaluation : (t : Tm Δ s) (σ : Sub Γ Δ) → E ⊢ Γ ▷ (⦅_⦆ {M = M Γ} t .apply σ) ≡ (t [ σ ])
evaluation (var x)    σ = refl (σ x)
evaluation (op • ts) σ = app (λ i → evaluation (ts i) σ)
```

The term model satisfies all the equations it started out with.

```
satisfies : ∀ i → M Γ ⊨ E i
satisfies i σ = begin
  ⦅ tₗ ⦆ .apply σ ≈⟨ evaluation tₗ σ ⟩
  tₗ [ σ ]        ≈⟨ sub (hyp i) σ ⟩
  tᵣ [ σ ]        ≈˘⟨ evaluation tᵣ σ ⟩
  ⦅ tᵣ ⦆ .apply σ ∎
  where
  open SetoidReasoning (TmSetoid _ _)
  tₗ = E i .Eq.lhs
  tᵣ = E i .Eq.rhs
```

### 7.2 Completeness

Birkhoff's completeness theorem [1935]: Any valid consequence is derivable in the equational theory.

```
module Completeness {I : Set ℓⁱ} (E : I → Eq) {Γ s} {t t′ : Tm Γ s} where
  open TermModel E

  completeness : E ⊃ (t ≐ t′) → E ⊢ Γ ▹ t ≡ t′
  completeness V = begin
    t              ≈˘⟨ identity t ⟩
    t [ σ₀ ]       ≈˘⟨ evaluation t σ₀ ⟩
    ⦅ t ⦆ .apply σ₀  ≈⟨ V (M Γ) satisfies σ₀ ⟩
    ⦅ t′ ⦆ .apply σ₀ ≈⟨ evaluation t′ σ₀ ⟩
    t′ [ σ₀ ]      ≈⟨ identity t′ ⟩
    t′             ∎
    where open SetoidReasoning (TmSetoid Γ s)
```

Q.E.D.

## 8 Related work

Gunther et al. [2017] further formalize signature morphisms. These would be, in our setting, morphisms of indexed containers, described by Altenkirch et al. [2015], albeit in a slightly different semantics, slice categories.

DeMeo's rather comprehensive development [2021] formalizes single-sorted algebras up to the Birkhoff's HSP theorem in Agda. DeMeo's signatures are containers; even though he does not make this connection explicit, it inspired the use of indexed containers in the present development. DeMeo's formalization is basis for `https://github.com/ualib/agda-algebras`.

Amato et al. [2021] formalize multi-sorted algebras with finitary operators in UniMath. Limiting to finitary operators is due to the restrictions of the UniMath type theory, which does not have W-types nor user-defined inductive types. These restrictions also prompt the authors to code terms as lists of stack machine instructions rather than trees.

Lynge and Spitters [2019] formalize multi-sorted algebras in HoTT, also restricting to finitary operators. Using HoTT they can define quotients as types, obsoleting setoids. They prove three isomorphism theorems concerning sub- and quotient algebras. A universal algebra or varieties are not formalized.

**Acknowledgments.** Standing on the shoulders of giants: the present concise formalization is enabled by the well-organized standard library of Agda maintained by Matthew Daggitt and Guillaume Allais. Thanks to Jacques Carette for discussions and constructive review of the Agda code.

This document has been generated from `MultiSortedAlgebra.agda` using the `agda2lagda` translator and the `agda --latex` backend.

## References

T. Altenkirch, N. Ghani, P. G. Hancock, C. McBride, and P. Morris. Indexed containers. *J. Func. Program.*, 25, 2015. doi: 10.1017/S095679681500009X. URL `https://doi.org/10.1017/S095679681500009X`.

G. Amato, M. Maggesi, and C. P. Brogi. Universal algebra in unimath. *CoRR*, abs/2102.05952, 2021. URL `https://arxiv.org/abs/2102.05952`.

G. Birkhoff. On the structure of abstract algebras. *Mathematical Proceedings of the Cambridge Philosophical Society*, 31(4):433–454, 1935. doi: 10.1017/S0305004100013463.

W. J. DeMeo. The Agda Universal Algebra Library and Birkhoff's theorem in Martin-Löf dependent type theory. *CoRR*, abs/2101.10166, 2021. URL `https://arxiv.org/abs/2101.10166`.

E. Gunther, A. Gadea, and M. Pagano. Formalization of universal algebra in Agda. In S. Alves and R. Wasserman, editors, *12th Workshop on Logical and Semantic Frameworks, with Applications, LSFA 2017, Brasília, Brazil, September 23-24, 2017*, volume 338 of *Electr. Notes in Theor. Comp. Sci.*, pages 147–166. Elsevier, 2017. doi: 10.1016/j.entcs.2018.10.010. URL `https://doi.org/10.1016/j.entcs.2018.10.010`.

A. Lynge and B. Spitters. Universal algebra in HoTT. In M. Bezem, editor, *TYPES 2019, 25th International Conference on Types for Proofs and Programs*, 2019. URL `http://www.ii.uib.no/~bezem/abstracts/TYPES_2019_paper_7`.